\documentclass[prb,showpacs,superscriptaddress,twocolumn]{revtex4}%
\usepackage{amsfonts}
\usepackage{amsmath}
\usepackage{amssymb}
\usepackage{graphicx}%
\begin{document}
\title{Spin Hall effect in the kagom\'{e} lattice with Rashba spin-orbit interaction}
\author{Guocai Liu}
\affiliation{Institute of Semiconductor, Chinese Academy of Sciences, Beijing 100083,
People's Republic of China}
\author{Ping Zhang}
\thanks{Author to whom correspondence should be addressed. Email
address: zhang\_ping@iapcm.ac.cn} \affiliation{LCP, Institute of
Applied Physics and Computational Mathematics, P.O. Box 8009,
Beijing 100088, People's Republic of China} \affiliation{Center for
Applied Physics and Technology, Peking University, Beijing 100871,
People's Republic of China}
\author{Zhigang Wang}
\affiliation{LCP, Institute of Applied Physics and Computational Mathematics, P.O. Box
8009, Beijing 100088, People's Republic of China}
\author{Shu-Shen Li}
\affiliation{Institute of Semiconductor, Chinese Academy of
Sciences, Beijing 100083, People's Republic of China}

\begin{abstract}
We study the spin Hall effect in the kagom\'{e} lattice with Rashba spin-orbit
coupling. The conserved spin Hall conductance $\sigma_{xy}^{s}$ (see text) and
its two components, i.e., the conventional term $\sigma_{xy}^{s0}$ and the
spin-torque-dipole term $\sigma_{xy}^{s\tau}$, are numerically calculated,
which show a series of plateaus as a function of the electron Fermi energy
$\epsilon_{F}$. A consistent two-band analysis, as well as a Berry-phase
interpretation, is also given. We show that these plateaus are a consequence
of the various Fermi-surface topologies when tuning $\epsilon_{F}$. In
particular, we predict that compared to the case with the Fermi surface
encircling the $\mathbf{\Gamma}$ point in the Brillouin zone, the amplitude of
the spin Hall conductance with the Fermi surface encircling the $\mathbf{K}$
points is twice enhanced, which makes it highly meaningful in the future to
systematically carry out studies of the $\mathbf{K}$-valley spintronics.

\end{abstract}
\maketitle

Spintronics, which combines the basic quantum mechanics of coherent spin
dynamics and technological applications in information processing and storage
devices, has been become a very active and promising field
\cite{Wolf,Awsch,Das}. The key is how to control and manipulate the spin
degrees of freedom. One of the tools is using the spin-orbit (SO) couplings,
which describe the interactions between the electron's orbital and spin
degrees and provide an ability to manipulate the spin state via changing some
external factors, such as an external electric field. It has been argued that
the SO interaction leads to an intrinsic spin Hall effect (SHE)
\cite{Muk1,Sinova}, in which a spin current flows perpendicular to an applied
electric field. The initial theoretical
\cite{Muk1,Sinova,Shen2004,Dimi2004,Ino,Rashba2004,Bern2004,Sch,Guo2005,Bern2005,Nic2005,Sun2005}
and experimental \cite{Kato,Wund,Sih,Sih2006} studies of SHE were mainly
focused on the $p$ or $n$ doped semiconductors (such as GaAs). Then, Murakami
\textit{et}. \textit{al}. \cite{Mura2004} first identified a class of cubic
materials that are usual insulators, but nonetheless exhibit a finite spin
Hall conductance (SHC). In those proposed \textquotedblleft spin Hall
insulators\textquotedblright\ (SHIs) the SHC is not quantized and depends on
the system parameters. Later and even more fundamentally, it has been evolving
into one important theme in condensed matter physics that the SHC can be
quantized in time-reversal invariant systems and thus can be used as an order
parameter to characterize the emergence of new topological insulating state of
matter
\cite{Kane1,Kane2,Onoda,Ber,Qi,Sheng,Fukui,Fu,Fu1,Fu2,Onoda2007,Wu,Xu,Ber2006,Moore}%
.

It is clear now that besides the external SO coupling (e.g., the Rashba SO
coupling), the lattice structure itself also has crucial impact on SHE through
the related band structure. Different lattice structure may produce new
features in the spin transport, which provides versatile choices of materials
to study spin Hall transport. Motivated by this observation, in this paper we
study the intrinsic SHE of the noninteracting electrons in a two-dimensional
(2D) kagom\'{e} lattice with Rashba SO coupling. Since our attention is solely
on the SHE character brought about by the interplay between the kagom\'{e}
lattice structure and the Rashba SO coupling, thus unlike most of previous
works, the kagom\'{e} lattice considered in this paper is nonmagnetic. The
nonmagnetic kagom\'{e} lattice structure has been either fabricated by modern
patterning techniques \cite{Mohan2003,Hig2000} or observed in reconstructed
semiconductor surfaces \cite{Tong1985}. In the former case, remarkably, the
electron filling factor (namely, the Fermi energy) can be readily controlled
by applying a gate voltage \cite{Shi2001}. Our lattice model is free from the
constraint imposed on the $\mathbf{k}\mathtt{\cdot}\mathbf{p}$ approximation
used in the extensively studied GaAs two-dimensional electron gas (2DEG), in
which the $\mathbf{k}\mathtt{\cdot}\mathbf{p}$ Hamiltonian is only valid
around the $\Gamma$ point in the Brillouin zone (BZ). In contrast, our lattice
model allows for any electron filling, which result in various Fermi-surface
topologies, which in turn, as will be shown below, produces profound effects
on the spin Hall transport.

To calculate the SHC and build a correspondence between spin current and spin
accumulation in the present SO coupled system, in which the electron spin
($s_{z}$ here, to be specific) is not conserved, we use a \textquotedblleft
conserved\textquotedblright\ spin current $\mathcal{J}_{s}$ \cite{Shi}, which
is a sum of the conventional spin current $\mathbf{J}_{s}\mathtt{\equiv}%
\frac{1}{2}\{\mathbf{v},s_{z}\}$ and a spin torque dipole $\mathbf{P}_{\tau
}\mathtt{\equiv}\mathbf{r}\dot{s}_{z}$. This spin current satisfies both spin
continuity equation $\partial_{t}s_{z}\mathtt{+}\nabla\mathtt{\cdot
}\mathcal{J}_{s}\mathtt{=}0$ (within spin relaxation time) and Onsager
relation \cite{Ping2004}. If the spin itself is conserved (as in quantum
SHIs), $\mathcal{J}_{s}$ is reduced to $\mathbf{J}_{s}$. In general, the spin
transport coefficient $\sigma_{\mu\nu}^{s}$ under new definition is composed
of two parts, i.e., the conventional part $\sigma_{\mu\nu}^{s0}$ and the spin
torque dipole correction $\sigma_{\mu\nu}^{s\tau}$. A general Kubo formula
\cite{Ping2005} for the spin transport coefficients is employed in this paper
to calculate the SHC.

Let us consider the tight-binding model for independent electrons on the 2D
kagom\'{e} lattice (Fig. 1). The spin-independent part of the Hamiltonian is
given by
\begin{equation}
H_{0}=t_{0}\sum_{\langle i,j\rangle}(c_{i\alpha}^{\dag}c_{j\alpha}%
+\text{H.c.}), \label{Hnn}%
\end{equation}
where $t_{ij}\mathtt{=}t_{0}$ is the hopping amplitude between the nearest
neighbor link $\langle i,j\rangle$, $c_{i\alpha}^{\dag}$ ($c_{i\alpha}$) is
the creation (annihilation) operator of an electron with spin $\alpha$ (up or
down) on lattice site $i$. For simplicity, we choose $t_{0}$=$1$ as the energy
unit and the distance $a$ between the nearest sites as the length unit
throughout this paper.

\begin{figure}[ptb]
\begin{center}
\includegraphics[width=0.6\linewidth]{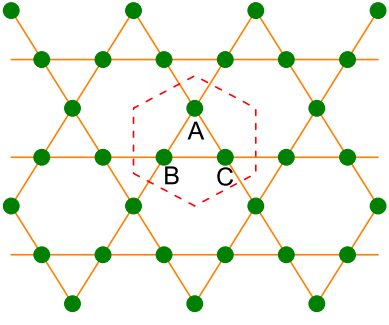}
\end{center}
\caption{(Color online) Schematic picture of the 2D kagom\'{e} lattice. The
dashed lines represent the Wigner-Seitz unit cell, which contains three
independent sites (A, B, C).}%
\end{figure}

The Hamiltonian (\ref{Hnn}) can be diagonalized in the momentum space as%
\begin{equation}
\mathcal{H}_{0}=\sum_{\mathbf{k}}\psi_{\mathbf{k}}^{+}(H_{0}(\mathbf{k}%
)\mathtt{\otimes}\mathbf{I}_{2\times2})\psi_{\mathbf{k}}, \label{Hk}%
\end{equation}
where the $2\mathtt{\times}2$ unit matrix $\mathbf{I}_{2\times2}$ denotes the
spin degeneracy in the Hamiltonian $H_{0}$, and $\psi_{\mathbf{k}}%
$=$(c_{A\mathbf{k}\uparrow},c_{B\mathbf{k}\uparrow},c_{C\mathbf{k}\uparrow
},c_{A\mathbf{k}\downarrow},c_{B\mathbf{k}\downarrow},c_{C\mathbf{k}%
\downarrow})^{\text{T}}$ is the six-component electron field operator, which
includes the three lattice sites $s$ ($\mathtt{=}A,B,C$) in the Wigner-Seitz
unit cell shown in Fig. 1. Each component of $\psi_{\mathbf{k}}$ is the
Fourier transform of $c_{i\alpha}$, i.e.,
\begin{equation}
\psi_{s\alpha}(\mathbf{k})=\sum_{mn}c_{mns\alpha}e^{i\mathbf{k\cdot r}_{mns}},
\label{Fourier}%
\end{equation}
where we have changed notation $i\mathtt{\rightarrow}(mns)$ by using $(mn)$ to
label the kagom\'{e} unit cells. $H_{0}(\mathbf{k})$ is a $3\mathtt{\times}3$
spinless matrix given by
\begin{equation}
H_{0}(\mathbf{k})=\left(
\begin{array}
[c]{ccc}%
0 & 2\cos\left(  \mathbf{k}\mathtt{\cdot}\mathbf{a}_{1}\right)  & 2\cos\left(
\mathbf{k}\mathtt{\cdot}\mathbf{a}_{3}\right) \\
2\cos\left(  \mathbf{k}\mathtt{\cdot}\mathbf{a}_{1}\right)  & 0 & 2\cos\left(
\mathbf{k}\mathtt{\cdot}\mathbf{a}_{2}\right) \\
2\cos\left(  \mathbf{k}\mathtt{\cdot}\mathbf{a}_{3}\right)  & 2\cos\left(
\mathbf{k}\mathtt{\cdot}\mathbf{a}_{2}\right)  & 0
\end{array}
\right)  , \label{Hamilton}%
\end{equation}
where $\mathbf{a}_{1}$=$(-1/2,-\sqrt{3}/2)$, $\mathbf{a}_{2}$=$(1,0)$, and
$\mathbf{a}_{3}$=$(-1/2,\sqrt{3}/2)$ represent the displacements in a unit
cell from A to B site, from B to C site, and from C to A site, respectively.
In this notation, the first BZ is a hexagon with the corners of $\mathbf{K}%
$=$\pm\left(  2\pi/3\right)  \mathbf{a}_{1}$, $\pm\left(  2\pi/3\right)
\mathbf{a}_{2}$, $\pm\left(  2\pi/3\right)  \mathbf{a}_{3}$.

The energy spectrum for spinless Hamiltonian $H_{0}(\mathbf{k})$ is
characterized by one dispersionless flat band ($\epsilon_{1\mathbf{k}}^{(0)}%
$=$-2$), which reflects the fact that the 2D kagom\'{e} lattice is a line
graph of the honeycomb structure \cite{Mie}, and two dispersive bands,
$\epsilon_{2(3)\mathbf{k}}^{(0)}=1\mathtt{\mp}\sqrt{4b_{\mathbf{k}}-3}$ with
$b_{\mathbf{k}}$=$\sum_{i=1}^{3}\cos^{2}\left(  \mathbf{k\cdot a}_{i}\right)
$. These two dispersive bands touch at the corners (\textbf{K}-points) of the
BZ and exhibit Dirac-type energy spectra, $\epsilon_{2(3)\mathbf{k}}%
^{(0)}\mathtt{=}(1\mathtt{\mp}\sqrt{3}|\mathbf{k}\mathtt{-}\mathbf{K}|)$,
which implies a \textquotedblleft particle-hole\textquotedblright\ symmetry
with respect to the Fermi energy $\epsilon_{F}\mathtt{=}1$. The corresponding
eigenstates of $H_{0}(\mathbf{k})$ are given by%
\begin{equation}
\left\vert u_{n\mathbf{k}}^{(0)}\right\rangle =G_{n\mathbf{k}}\left(
q_{1\mathbf{k}},q_{2\mathbf{k}},q_{3\mathbf{k}}\right)  ^{\text{T}},
\end{equation}
where the expressions of the components $q_{i\mathbf{k}}$ and the normalized
factor $G_{n}(\mathbf{k})$ for each band are given in Table I. At two
equivalent BZ edge points $\mathbf{M}$=$(0,\pm\pi/\sqrt{3})$, one can find
that the wave function $|u_{n\mathbf{k}}^{(0)}\rangle$ is ill defined since
both its denominator and numerator are zero at these two points.

When an external Rashba SO coupling, which can be realized by a perpendicular
electric field or by interaction with a substrate, is taken into account in
the 2D kagom\'{e} lattice model, the spin degeneracy will be lifted. The
tight-binding expression for this external Rashba term can be given as
follows
\begin{equation}
H_{\text{SO}}=i\frac{\lambda}{\hslash}\sum_{\langle ij\rangle\alpha\beta
}c_{i\alpha}^{\dag}(\mathbf{\sigma\times\hat{d}}_{ij})_{z}c_{j\beta},
\end{equation}
where $\lambda$ is the Rashba coefficient,\textbf{ }$\mathbf{\sigma}$ are the
Pauli matrices and $\mathbf{\hat{d}}_{ij}$ is a vector along the bond the
electron traverses going from site $j$ to $i$. Taking the Fourier transform
[Eq. (\ref{Fourier})] and considering the $\psi_{\mathbf{k}}$ below Eq.
(\ref{Hk}), we have $H_{\text{SO}}\mathtt{=}\sum_{\mathbf{k}}\psi_{\mathbf{k}%
}^{+}H_{\text{SO}}(\mathbf{k})\psi_{\mathbf{k}}$ with
\begin{equation}
H_{\text{SO}}(\mathbf{k})=\left(
\begin{array}
[c]{cc}%
0 & H_{R}(\mathbf{k})\\
H_{R}^{\ast}(\mathbf{k}) & 0
\end{array}
\right)
\end{equation}
and%
\begin{equation}
H_{R}(\mathbf{k})=\lambda\left(
\begin{array}
[c]{ccc}%
0 & e^{i\frac{\pi}{6}}\sin(\mathbf{k\mathtt{\cdot}a}_{1}) & -e^{-i\frac{\pi
}{6}}\sin(\mathbf{k\mathtt{\cdot}a}_{3})\\
e^{i\frac{\pi}{6}}\sin(\mathbf{k\mathtt{\cdot}a}_{1}) & 0 & -i\sin
(\mathbf{k\mathtt{\cdot}a}_{2})\\
-e^{-i\frac{\pi}{6}}\sin(\mathbf{k\mathtt{\cdot}a}_{3}) & -i\sin
(\mathbf{k\mathtt{\cdot}a}_{2}) & 0
\end{array}
\right)  . \label{HR}%
\end{equation}
Inclusion of the Rashba SO term in the Hamiltonian makes the analytical
derivation of the eigenstates $\left\vert u_{n\mathbf{k}}\right\rangle $
($n\mathtt{=}1,...,6$) and eigenenergies $\epsilon_{n\mathbf{k}}$ very
tedious. At the general $k$ points, these quantities can only be numerically
obtained. At some high-symmetry $k$ points, however, they can be approximately
obtained, which turns out to provide a great help in analyzing SHC.

The energy spectrum for the total Hamiltonian $H(\mathbf{k})\mathtt{=}%
H_{0}(\mathbf{k})\mathtt{+}H_{\text{SO}}(\mathbf{k})$ is numerically
calculated and shown in Fig. 2 (solid curves) along the high-symmetry lines
($\mathbf{\Gamma}\mathtt{\rightarrow}\mathbf{K}$, $\mathbf{K}%
\mathtt{\rightarrow}\mathbf{M}$, and $\mathbf{M}\mathtt{\rightarrow
}\mathbf{\Gamma}$) in the BZ. The Rashba coefficient is chosen to be $\lambda
$=$0$.$1$. Note that in this paper, we only concern the physically reasonable
limit of $\lambda\mathtt{\ll}t_{0}$ ($t_{0}$ is chosen to be unity). For
comparison we also plot in Fig. 2 (dashed curves) the energy spectrum in the
absence of the Rashba SO coupling ($\lambda$=$0$). For the middle and upper
bands, one can see that the spin degeneracies are generally lifted in the BZ
with the exception at $\mathbf{\Gamma}$ and $\mathbf{M}$ points, at which the
energy is still spin degenerate due to time-reversal symmetry. The most
prominent splitting occurs at the corners ($\mathbf{K}$-points) of the BZ.
However, this splitting does not change the Dirac-type nature of the
dispersions around these corners. Also, there still exists the contacts at
these corners between one middle band and one upper band, as seen from Fig. 2.
For the lowest flat band, on the other hand, it reveals in Fig. 2 that the
Rashba splitting is negligibly small, and there is no observable SO effect on
this flat band. The two-band approximation given below will also indicate this
fact. \begin{table}[th]
\caption{The expressions for the coefficients in Eq. (5) with $x_{i}%
$=$\mathbf{k\mathtt{\cdot}a}_{i}$.}
\begin{tabular}
[c]{cc}\hline\hline
$q_{1k}$ & $\ \ \ \ \ \ \ \ \ \ \ \ \ \ \ \ \ \ \frac{1}{2}[\epsilon
_{nk}^{(0)2}-4\cos^{2}x_{2}]$\\
$q_{2k}$ & $\ \ \ \ \ \ \ \ \ \ \ \ \ \ \ \ \ \ \epsilon_{nk}^{(0)}\cos
x_{1}+2\cos x_{2}\cos x_{3}$\\
$q_{3k}$ & $\ \ \ \ \ \ \ \ \ \ \ \ \ \ \ \ \ \ \epsilon_{nk}^{(0)}\cos
x_{3}+2\cos x_{2}\cos x_{1}$\\
$G_{nk}^{-2}$ & $\ \ \ \ \ 2b_{k}\epsilon_{nk}^{(0)2}+[4b_{k}-3\epsilon
_{nk}^{(0)2}]\cos^{2}x_{2}+6(b_{k}-1)\epsilon_{nk}^{(0)}\ \ $\\\hline
\end{tabular}
\end{table}

The conserved SHC $\sigma_{xy}^{s}$ includes two components, $\sigma_{xy}%
^{s}\mathtt{=}\sigma_{xy}^{s0}\mathtt{+}\sigma_{xy}^{s\tau}$, where
$\sigma_{xy}^{s0}$ is the conventional part and $\sigma_{xy}^{s\tau}$ comes
from the spin torque dipole correction. In terms of the band energies
$\epsilon_{n\mathbf{k}}$ and states $|u_{n\mathbf{k}}\rangle$ of
$H(\mathbf{k})\mathtt{=}H_{0}(\mathbf{k})\mathtt{+}H_{\text{SO}}(\mathbf{k})$,
these two SHC components are given by \cite{Ping2005}
\begin{align}
\sigma_{xy}^{s0}  &  =-e\hslash\sum_{n\neq n^{\prime},\mathbf{k}}\left[
f(\epsilon_{n\mathbf{k}})-f(\epsilon_{n^{\prime}\mathbf{k}})\right]
\nonumber\\
&  \times\frac{\operatorname{Im}\langle u_{n\mathbf{k}}|\frac{1}{2}\left\{
v_{x},s_{z}\right\}  |u_{n^{\prime}\mathbf{k}}\rangle\langle u_{n^{\prime
}\mathbf{k}}|v_{y}|u_{n\mathbf{k}}\rangle}{\left(  \epsilon_{n\mathbf{k}%
}-\epsilon_{n^{\prime}\mathbf{k}}\right)  ^{2}+\eta^{2}} \label{Conventional}%
\end{align}
and
\begin{align}
\sigma_{xy}^{s\tau}  &  =-e\hslash\lim_{\mathbf{q}\rightarrow0}\frac{1}{q_{x}%
}\sum_{n\neq n^{\prime},\mathbf{k}}\left[  f(\epsilon_{n\mathbf{k}%
})-f(\epsilon_{n^{\prime}\mathbf{k+q}})\right] \nonumber\\
&  \times\frac{\operatorname{Re}\langle u_{n\mathbf{k}}|\tau\left(
\mathbf{k},\mathbf{q}\right)  |u_{n^{\prime}\mathbf{k+q}}\rangle\langle
u_{n^{\prime}\mathbf{k+q}}|v_{y}\left(  \mathbf{k},\mathbf{q}\right)
|u_{n\mathbf{k}}\rangle}{\left(  \epsilon_{n\mathbf{k}}-\epsilon_{n^{\prime
}\mathbf{k+q}}\right)  ^{2}+\eta^{2}}, \label{Torque}%
\end{align}
where $\tau\left(  \mathbf{k},\mathbf{q}\right)  \mathtt{\equiv}\frac{1}%
{2}\left[  \tau\left(  \mathbf{k}\right)  \mathtt{+}\tau\left(  \mathbf{k}%
+\mathbf{q}\right)  \right]  $ with $\tau\left(  \mathbf{k}\right)
\mathtt{=}\dot{s}_{z}$, $\mathbf{v}\left(  \mathbf{k},\mathbf{q}\right)  $ is
given in the same manner, and $f(\epsilon_{n\mathbf{k}})$ is the equilibrium
Fermi function. The limit of $\eta\rightarrow0$ should be taken at the last
step of the calculation. In the present six-band model the spin operator
$s_{z}$ should be written as $\mathbf{I}_{3\times3}\mathtt{\otimes}\sigma_{z}$
in unit of $\hslash/2$. \begin{figure}[ptb]
\begin{center}
\includegraphics[width=1.0\linewidth]{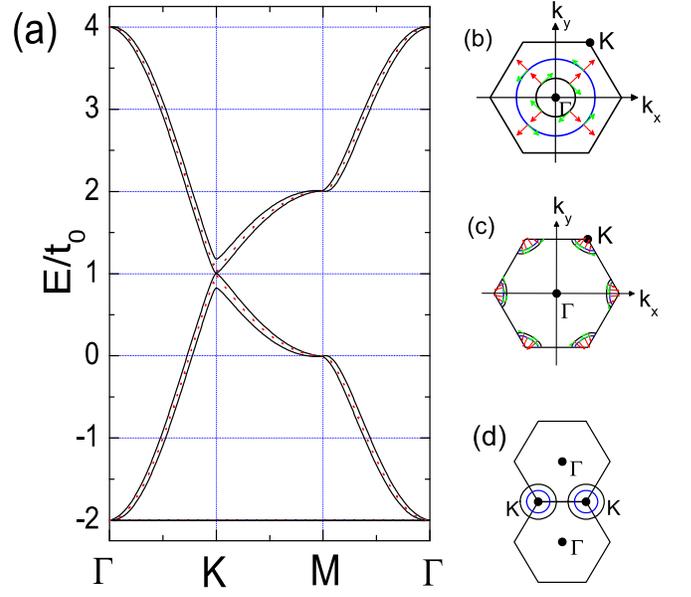}
\end{center}
\caption{(Color online) (a) Energy spectrum of the 2D kagom\'{e} lattice with
Rashba SO constant $\lambda$=$0.1$ (solid curves). (b) and (c) show the Fermi
surfaces in the regimes $-$2$<$$\epsilon_{F}\mathtt{<}0$ and 0$<$$\epsilon
_{F}\mathtt{<}1$, respectively. Directions of the electron's velocity and spin
polarization are also shown by red and green arrows respectively. (d)
Reconstructed Fermi surface around the two \textbf{K} points by gluing the six
sheets of the Fermi surface in (c). For comparison, the energy spectrum in the
absence of the Rashba SO\ coupling is also plotted, see the dashed curves in
(a). One can see that the lowest flat band is immune to the Rashba SO
coupling. }%
\end{figure}

We have numerically calculated the SHC as a function of the electron Fermi
energy $\epsilon_{F}$. The main results for zero temperature are shown in Fig.
3, in which Fig. 3(a) plots the conserved SHC $\sigma_{xy}^{s}$, while Fig.
3(b) plots its two components, i.e., the conventional term $\sigma_{xy}^{s0}$
and the spin torque dipole term $\sigma_{xy}^{s\tau}$. For comparison, the
value of the Rashba SO coefficient $\lambda$ used in Fig. 3 is the same as in
Fig. 2 (solid curves). One can see that within the whole range of the electron
filling (Fermi energy), the two components $\sigma_{xy}^{s0}$ and $\sigma
_{xy}^{s\tau}$ always oppose each other. In fact, this feature of opposite
signs of the two components $\sigma_{xy}^{s0}$ and $\sigma_{xy}^{s\tau}$ (if
both of them are nonzero) is robust and does not depend on specific models
\cite{Note}. Remarkably, the amplitude of $\sigma_{xy}^{s\tau}$ is as twice
large as that of $\sigma_{xy}^{s0}$, which results in the consequence that the
total SHC $\sigma_{xy}^{s}$ has an overall sign change with respect to the
conventional SHC $\sigma_{xy}^{s0}$. Together with the previous studies of the
conserved SHC in the Rashba 2DEG \cite{Ping2005}. As will be shown below,
around the $\mathbf{\Gamma}$ point the present model can be mapped into the
simple Rashba 2DEG model. Thus, one can see the key role played by the
spin-torque-dipole term, which in some special cases tends to overwhelm the
conventional SHC by an opposite contribution. On the other hand, considering
the variation of the SHC as a function of electron Fermi energy, the present
results in the 2D kagom\'{e} lattice display more profound features compared
to those in the 2DEG system. In fact, it reveals in Fig. 3 that the conserved
SHC and its two components display four plateaus as a function $\epsilon_{F}$.
When the electron filling satisfies the condition $-$2.0$\mathtt{<}%
\epsilon_{F}\mathtt{<}0$, the value of $\sigma_{xy}^{s}$ is $e/8\pi$, while
the values of $\sigma_{xy}^{s0}$ and $\sigma_{xy}^{s\tau}$ are $-e/8\pi$ and
$e/4\pi$, respectively. When the electron filling increases to satisfy
$0\mathtt{<}\epsilon_{F}\mathtt{<}$1.0, then the conserved SHC jumps down to
$\sigma_{xy}^{s}\mathtt{=-}e/4\pi$, while its two components also jump to
$\sigma_{xy}^{s0}\mathtt{=}e/4\pi$ and $\sigma_{xy}^{s\tau}$=$\mathtt{-}%
e/2\pi$. When the Fermi energy continues to increase to satisfy 1.0$\mathtt{<}%
\epsilon_{F}\mathtt{<}$2.0, then the conserved SHC jumps up to $\sigma
_{xy}^{s}\mathtt{=}e/4\pi$, while its two components also jump to $\sigma
_{xy}^{s0}\mathtt{=-}e/4\pi$ and $\sigma_{xy}^{s\tau}\mathtt{=}e/2\pi$.
Finally, when the Fermi energy satisfies the condition 2.0$\mathtt{<}%
\epsilon_{F}\mathtt{<}$4.0, then the conserved SHC jumps down to $\sigma
_{xy}^{s}\mathtt{=-}e/8\pi$, while its two components jump to $\sigma
_{xy}^{s0}\mathtt{=}e/8\pi$ and $\sigma_{xy}^{s\tau}$\texttt{=}$\mathtt{-}%
e/4\pi$.

\begin{figure}[ptb]
\begin{center}
\includegraphics[width=1.0\linewidth]{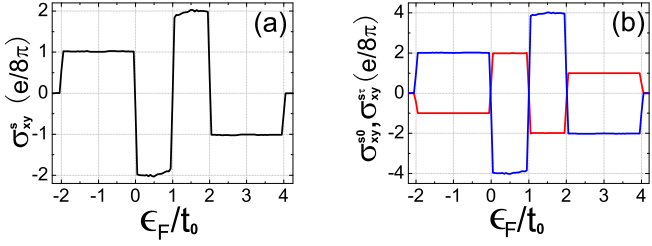}
\end{center}
\caption{(Color online) (a) The conserved SHC $\sigma_{xy}^{s}$ and (b) its
two components $\sigma_{xy}^{0}$ (red curve) and $\sigma_{xy}^{\tau}$ (blue
curve) as a function of the electron Fermi energy for the Rashba coefficient
$\lambda$=$0.1$. }%
\end{figure}

We turn now to understand the physics embodied in Fig. 3. Since we are dealing
with the usual case of weak SO coupling ($\lambda\mathtt{\ll}t_{0}$), thus the
SHC behavior in Fig. 3 should be mainly due to the coupling of the two Rashba
SO-split bands and can be described by an effective two-band approximation. To
be more clear, let us treat the Rashba SO term as a perturbation to the
spinless Hamiltonian $H_{0}(\mathbf{k})$. The expressions for the unperturbed
eigenenergies $\epsilon_{n\mathbf{k}}^{(0)}$ ($n\mathtt{=}1,2,3$) and
eigenstates $|u_{n\mathbf{k}}^{(0)}\rangle$ have been given above. Then, the
effective two-band Hamiltonian originating from $\epsilon_{n\mathbf{k}}^{(0)}$
and $|u_{n\mathbf{k}}^{(0)}\rangle$ is obtained by taking into account the
Rashba SO splitting as follows%

\begin{equation}
\bar{H}_{n}(\mathbf{k})=\epsilon_{n\mathbf{k}}^{(0)}\mathbf{I}_{2\times
2}+\left(
\begin{array}
[c]{cc}%
0 & \Delta_{n\mathbf{k}}e^{i\varphi_{n\mathbf{k}}}\\
\Delta_{n\mathbf{k}}e^{-i\varphi_{n\mathbf{k}}} & 0
\end{array}
\right)  , \label{H2}%
\end{equation}
where the basis set to expand $\bar{H}_{n}(\mathbf{k})$ consists of
$|u_{n\mathbf{k}}^{(0)}\rangle\mathtt{\otimes}|\uparrow\rangle$ and
$|u_{n\mathbf{k}}^{(0)}\rangle\mathtt{\otimes}|\downarrow\rangle$. Here the
coefficients $\Delta_{n\mathbf{k}}$ and $\varphi_{n\mathbf{k}}$ are defined
by
\begin{widetext}
\begin{align}
\Delta_{n\mathbf{k}}\cos\varphi_{n\mathbf{k}}  &  \text{=}\mathtt{-}%
\frac{\sqrt{3}\lambda}{2}G_{n}^{2}(\mathbf{k})(\epsilon_{n\mathbf{k}}%
^{(0)}\mathtt{+}2)(\epsilon_{n\mathbf{k}}^{(0)2}\mathtt{-}4\cos^{2}k_{x})\cos
k_{x}\sin(\sqrt{3}k_{y}),\label{deta}\\
\Delta_{n\mathbf{k}}\sin\varphi_{n\mathbf{k}}  &  \text{=}\mathtt{-}%
\frac{\lambda}{2}G_{n}^{2}(\mathbf{k})(\epsilon_{n\mathbf{k}}^{(0)}%
\mathtt{+}2)\sin k_{x}\left[  4\epsilon_{n\mathbf{k}}^{(0)}\cos k_{x}%
\text{+}(\epsilon_{n\mathbf{k}}^{(0)2}\text{+}4\cos^{2}k_{x})\cos(\sqrt
{3}k_{y})\right]  .\nonumber
\end{align}
\end{widetext}
The eigenenergies of $\bar{H}_{n}(\mathbf{k})$ are
\begin{equation}
\epsilon_{n\mathbf{k}}^{(\pm)}=\epsilon_{n\mathbf{k}}^{(0)}\pm\Delta
_{n\mathbf{k}}.
\end{equation}
The corresponding eigenstates are given by
\begin{equation}
|u_{n\mathbf{k}}^{(\pm)}\rangle=\frac{1}{\sqrt{2}}\left(  \pm e^{i\varphi
_{n\mathbf{k}}},1\right)  ^{\text{T}}.
\end{equation}
As a result, the total Hamiltonian can now be approximated by%
\begin{equation}
\bar{H}(\mathbf{k})=\oplus_{n=1}^{3}\bar{H}_{n}(\mathbf{k})\text{.}
\label{add1}%
\end{equation}

This two-band approximation proves to work very well in the weak Rashba
SO\ coupling limit. In particular, one can see that the lowest\ flat band
($\epsilon_{1\mathbf{k}}^{(0)}\mathtt{=-}2)$ is not split by the Rashba SO
coupling in the first order in $\lambda$, since the quantity $\Delta
_{1\mathbf{k}}e^{i\varphi_{1\mathbf{k}}}$ is zero and as a result, the
off-diagonal element in Eq. (\ref{H2}) ($n\mathtt{=}1$) vanishes. This
perturbative analysis agrees well with the exact numerical result in Fig. 2,
which shows that the original flat band $\epsilon_{1\mathbf{k}}^{(0)}$ keeps
nearly dispersionless upon weak Rashba SO interaction. As a result, the
contribution of these two spin almost-degenerate flat bands to the SHC should
be negligibly small, which has been verified by our numerical test.

Thus, the finite SHC in Fig. 3 is ascribed to the contributions from the two
(SO-split) middle or the two upper bands, depending on the position of the
Fermi energy. Remarkably, there is a particle-hole symmetry between the middle
and upper bands with respect to their contact energy plane. As a consequence,
the SHC is antisymmetric with respect to the Fermi energy $\epsilon_{F}$=1.0,
as revealed in Fig. 3. Keeping this fact in mind, our remaining discussion of
Fig. 3 will focus on the two SHC plateaus and the transition between them when
scanning $\epsilon_{F}$ through the middle bands. According to Eqs.
(\ref{Conventional})-(\ref{Torque}) and our two-band approximation (\ref{H2}),
when the Fermi energy crosses the two middle bands $\epsilon_{2\mathbf{k}%
}^{(\pm)}$, it can be shown that the conventional part $\sigma_{xy}^{s0}$ and
the spin-torque-dipole part $\sigma_{xy}^{s\tau}$ of the conserved SHC are
given by
\begin{equation}
\sigma_{xy}^{s0}\text{=}\frac{e}{4}\sum_{\mathbf{k}}\frac{f_{2\mathbf{k}%
}^{(-)}\mathtt{-}f_{2\mathbf{k}}^{(+)}}{\Delta_{2\mathbf{k}}}\frac
{\partial\epsilon_{2\mathbf{k}}^{(0)}}{\partial k_{x}}\frac{\partial
\varphi_{2\mathbf{k}}}{\partial k_{y}} \label{es0}%
\end{equation}
and%
\begin{align}
\sigma_{xy}^{s\tau}  &  \text{=}\frac{e}{4}\sum_{\mathbf{k}}\frac
{f_{2\mathbf{k}}^{(-)}\mathtt{-}f_{2\mathbf{k}}^{(+)}}{\Delta_{2\mathbf{k}}%
}\left(  \frac{\partial\varphi}{\partial k_{x}}\frac{\partial\epsilon
_{2\mathbf{k}}^{(0)}}{\partial k_{y}}\mathtt{-}2\frac{\partial\varphi
_{2\mathbf{k}}}{\partial k_{y}}\frac{\partial\epsilon_{2\mathbf{k}}^{(0)}%
}{\partial k_{x}}\right) \nonumber\\
&  \mathtt{-}\frac{e}{4}\sum_{\mathbf{k}}\left(  \frac{\partial f_{2\mathbf{k}%
}^{(-)}}{\partial k_{x}}\text{+}\frac{\partial f_{2\mathbf{k}}^{(+)}}{\partial
k_{x}}\right)  \frac{\partial\varphi_{2\mathbf{k}}}{\partial k_{y}},
\label{est}%
\end{align}
where $f_{2\mathbf{k}}^{(\pm)}$ are the Fermi distribution functions for the
middle bands $\epsilon_{2\mathbf{k}}^{(\pm)}$.

According to the Kubo formulae (\ref{es0})-(\ref{est}), now let us see the
first SHC\ plateau in Fig. 3 for $-$2.0$\mathtt{<}\epsilon_{F}\mathtt{<}0$.
Since this plateau occurs upon occupation of the bottom (at the
$\mathbf{\Gamma}$ point) of the middle bands, thus we can simplify the
discussion of the first SHC\ plateau by expanding the middle-band Hamiltonian
$\bar{H}_{2}(\mathbf{k})$ around the $\mathbf{\Gamma}$ point up to the first
order in the Rashba coefficient $\lambda$
\begin{equation}
\bar{H}_{2}^{\mathbf{\Gamma}}=-2.0+k^{2}+\lambda\left(  k_{y}\sigma_{x}%
-k_{x}\sigma_{y}\right)  . \label{Gamma}%
\end{equation}
Not surprisingly, the effective Rashba Hamiltonian (\ref{Gamma}) around the
$\mathbf{\Gamma}$ point in the present kagom\'{e} lattice is similar to that
in the semiconductor 2DEG. Thus, as has been done in the 2DEG system
\cite{Ping2005}, a straightforward analytical calculation in terms of Eqs.
(\ref{es0})-(\ref{Gamma}) gives the zero-temperature SHC as $\sigma_{xy}%
^{s0}\mathtt{=-}e/8\pi$, $\sigma_{xy}^{s\tau}\mathtt{=}e/4\pi$, and
subsequently $\sigma_{xy}^{s}\mathtt{=}e/8\pi$. This analytical result is
consistent with the numerical result in Fig. 3 for the first SHC plateau.
Actually, the first SHC plateau in Fig. 3 goes beyond this analytical
treatment around the $\mathbf{\Gamma}$ point and persists with increasing the
Fermi energy up to $\epsilon_{F}$=$0$. The reason is attributed to the
equivalent Fermi-surface topologies when changing $\epsilon_{F}$ within the
interval $[-$2.0$,0]$. In fact, when $\epsilon_{F}$ lie in the region
$[-$2.0$,0]$, the corresponding 2D Fermi surface consists of of two simple
closed loops circling around the $\mathbf{\Gamma}$ point, as illustrated in
Fig. 2(b). Here, from Fig. 2(a) one can see that the critical value
$\epsilon_{F}$=$0$ corresponds to the case that the Fermi surface nested in
the middle bands touches the BZ edge at the $\mathbf{M}$ point, at which the
energies of the two middle band are degenerate due to the time-reversal symmetry.

When the Fermi level goes over this critical value, i.e., $\epsilon_{F}$%
$>$%
$0$, then the Fermi surface abruptly changes its topology. Instead of simple
closed loops, the Fermi surface for $0\mathtt{<}\epsilon_{F}\mathtt{<}1$.$0$
is characterized by six pieces of disconnected segments around six corners
($\mathbf{K}$ points$\mathbf{)}$ of the BZ as shown in Fig. 2(c). After gluing
these segments together by a simple translation operation in the extended BZ,
which does not change the property of electron states, then one can get two
sets of closed loops around two $\mathbf{K}$ points as shown in Fig. 2(d).
Thus the number of Fermi loops is doubled in the case of $0\mathtt{<}%
\epsilon_{F}\mathtt{<}1$.$0$ compared to the case of $-2.0\mathtt{<}%
\epsilon_{F}\mathtt{<}0$. This fundamental change in the Fermi-surface
topology by increasing the electron filling, together with the combined fact
that (i) the contributions from these two sets of $\mathbf{K}$-centered Fermi
loops are equivalent, and (ii) the normal direction of the Fermi surface for
$0\mathtt{<}\epsilon_{F}\mathtt{<}1$.$0$ is opposite to that for
$-2.0\mathtt{<}\epsilon_{F}\mathtt{<}0$, result in a downward jump of SHC
plateau from $\sigma_{xy}^{s}\mathtt{=}e/8\pi$ to $\sigma_{xy}^{s}%
\mathtt{=-}e/4\pi$ at the critical value of $\epsilon_{F}$=$0$. To be more
clear and to verify this argument based on the Fermi-surface topology, near
each corner of the BZ let us expand the middle-band Hamiltonian $\bar{H}%
_{2}(\mathbf{k})$ up to the first order in the Rashba coefficient $\lambda$,
\begin{equation}
\bar{H}_{2}^{\mathbf{K}}=1-\sqrt{3}k-\lambda\frac{\sqrt{3}}{2k}\left(
k_{y}\sigma_{x}-k_{x}\sigma_{y}\right)  , \label{K}%
\end{equation}
where the wave vector $\mathbf{k}$ is coordinated with respect to the
$\mathbf{K}$ point. By substitution of the eigenenergies and eigenstates of
$\bar{H}_{2}^{\mathbf{K}}$ into Eqs. (\ref{es0})-(\ref{est}), and taking into
account the six corners of the BZ, it is straightforward to obtain the
zero-temperature SHC as $\sigma_{xy}^{s0}\mathtt{=}e/4\pi$, $\sigma
_{xy}^{s\tau}\mathtt{=-}e/2\pi$, and $\sigma_{xy}^{s}\mathtt{=-}e/4\pi$, which
is consistent with the numerical result in Fig. 3.

Therefore, it becomes now clear that the different SHC plateaus in Fig. 3 are
due to the different Fermi-surface topologies when varying $\epsilon_{F}$.
This observation makes it highly interesting to reinterpret the
\textit{metallic} SHE, like what has been done in discussing the
\textit{metallic }AHE \cite{Jung,Fang,Yao,Haldane}, in terms of Berry phases
accumulated by adiabatic motion of electrons on the Fermi surface. The
previous work have shown the relationship between the SHC and the Berry phase
in the Rashba 2DEG \cite{Shen2004,Tsung}. The Fermi surface involved in those
discussions is as simple as shown in Fig. 2(b). Compared to the Rashba 2DEG,
one can see from the above discussions that the present kagom\'{e} lattice
provides more profound Fermi-surface topologies in the different regions of
the electron filling. On one hand, in the regime $-$2.0$\mathtt{<}\epsilon
_{F}\mathtt{<}$0 the effective \textquotedblleft$\Gamma$%
-valley\textquotedblright\ Hamiltonian (\ref{Gamma}) and the Fermi surface of
the kagom\'{e} lattice are identical to those of the Rashba 2DEG. As a result,
the two kinds of systems have the same Berry-phase SHC in this regime. On the
other hand, in the regime $0\mathtt{<}\epsilon_{F}\mathtt{<}1$.$0$ the
effective \textquotedblleft$\mathbf{K}$-valley\textquotedblright\ Hamiltonian
(\ref{K}) of the kagom\'{e} lattice, which is absent in the Rashba 2DEG, has a
remarkable Dirac-type spectrum with linear dependence of the energy on the
electron momentum. Exploring the $\mathbf{K}$-valley spintronics associated
with Berry phases is the task of our following discussions.

The Berry phases of Bloch states $|u_{n\mathbf{k}}^{(\pm)}\rangle$ for closed
paths $C_{n}^{(\pm)}$ in the $k$-space are written as
\begin{equation}
\gamma_{n}^{(\pm)}=\oint_{\mathbf{C}_{n}^{(\pm)}}\mathbf{A}_{n\mathbf{k}%
}^{(\pm)}\mathtt{\cdot}d\mathbf{k},
\end{equation}
where $\mathbf{C}_{n}^{(\pm)}$ are the Fermi loops identified by the
zero-temperature Fermi distribution function $\Theta(\epsilon_{F}%
-\epsilon_{n\mathbf{k}}^{\text{(}\pm\text{)}})$, and
\begin{equation}
\mathbf{A}_{n\mathbf{k}}^{(\pm)}\mathtt{=}\left\langle u_{n\mathbf{k}}^{(\pm
)}\right\vert (-i\frac{\partial}{\partial\mathbf{k}})\left\vert u_{n\mathbf{k}%
}^{(\pm)}\right\rangle \label{add3}%
\end{equation}
are the Berry connections. The corresponding Berry curvatures are defined as
$\mathbf{\Omega}_{n\mathbf{k}}^{(\pm)}\mathtt{=}\nabla_{\mathbf{k}%
}\mathtt{\times}\mathbf{A}_{n\mathbf{k}}^{(\pm)}$. By substituting Eq.
(\ref{add3}) into Eq. (\ref{Conventional}), and noting that $\sigma_{xy}%
^{s0}\mathtt{=-}\sigma_{yx}^{s0}$, we have%
\begin{equation}
\sigma_{xy}^{s0}\text{=}\mathtt{-}\frac{e\hbar}{2}\sum_{\mu=+,-}%
\sum_{\mathbf{k}}\frac{f_{n\mathbf{k}}^{(\mu)}}{\epsilon_{n\mathbf{k}}^{(\mu
)}-\epsilon_{n\mathbf{k}}^{(-\mu)}}\left[  \mathbf{v}_{n\mathbf{k}}%
^{(0)}\mathtt{\times}\mathbf{A}_{n\mathbf{k}}^{(\mu)}\right]  _{z},
\label{es1}%
\end{equation}
where $\mathbf{v}_{n\mathbf{k}}^{(0)}\mathtt{=}\frac{1}{\hbar}\frac
{\partial\epsilon_{n\mathbf{k}}^{(0)}}{\partial\mathbf{k}}$ is the band
velocity in the absence of the Rashba SO coupling. Now we focus our attention
to the regime 0$\mathtt{<}\epsilon_{F}\mathtt{<}$1.0, within which the gluing
Fermi surface consists of two set of loops around two \textbf{K} points as
shown in Fig. 2(d). According to the $\mathbf{K}$-valley\ Hamiltonian
(\ref{K}) and its eigenenergies $\epsilon_{2\mathbf{k}}^{\text{(}\pm\text{)}}%
$=$1\mathtt{-}\sqrt{3}k\pm\sqrt{3}\lambda/2$ and eigenstates $\left\vert
u_{2\mathbf{k}}^{(\pm)}\right\rangle $=$\frac{1}{\sqrt{2}}\left(  \mp
ie^{-i\varphi_{2\mathbf{k}}},1\right)  ^{\text{T}}$ with $\varphi
_{2\mathbf{k}}$= $\tan^{-1}(k_{y}/k_{x})$, it is straightforward to obtain the
zero-temperature conventional SHC as
\begin{equation}
\sigma_{xy}^{s0}\text{=}2\frac{e}{8\pi^{2}\lambda}\sum_{\mu=+,-}\mu
\int_{\mathbf{S}_{2}^{(\mu)}}d^{2}\mathbf{k}\left[  \frac{\mathbf{k}}{k}%
\times\mathbf{A}_{2\mathbf{k}}\right]  _{z}, \label{e23}%
\end{equation}
where $\mathbf{S}_{2}^{(\mu)}$ ($\mu$=$+,-$) in Eq. (\ref{e23}) denotes the
integral area bounded by the Fermi loops $C_{2}^{(\mu)}$ [see Fig. 2(d)], and
the Berry connections
\begin{equation}
\mathbf{A}_{2\mathbf{k}}^{(\pm)}\text{\texttt{=}}\mathtt{-}\frac{1}{2}%
\frac{\partial\varphi_{2\mathbf{k}}}{\partial\mathbf{k}}\text{=}\frac{1}%
{2}\left(  \frac{k_{y}}{k^{2}}\text{,}\mathtt{-}\frac{k_{x}}{k^{2}}\right)
\equiv\mathbf{A}_{2\mathbf{k}} \label{e24}%
\end{equation}
are equivalent for the two middle bands. Note that the factor 2 in Eq.
(\ref{e23}) is due to the contributions from the two \textbf{K} valleys.
Clearly, if we define an Abelian spin gauge field $\mathcal{B}_{2\mathbf{k}}%
$\textbf{=}$\left(  0,0,B\right)  $ with $B$=$\left[  \frac{\mathbf{k}}%
{k}\mathtt{\times}\mathbf{A}_{2\mathbf{k}}\right]  _{z}$, then Eq. (\ref{e23})
denotes a spin-flux difference through two areas $\mathbf{S}_{2}^{(+)}$ and
$\mathbf{S}_{2}^{(-)}$. In virtue of this way, we define a spin gauge
potential $\mathcal{A}_{2\mathbf{k}}$ to satisfy $\nabla_{\mathbf{k}%
}\mathtt{\times}\mathcal{A}_{2\mathbf{k}}$=$\mathcal{B}_{2\mathbf{k}}$, then
the expression (\ref{e23}) for the conventional SHC is rewritten as
\begin{align}
\sigma_{xy}^{s0}  &  \text{=}\frac{e}{4\pi^{2}\lambda}\sum_{\mu=+,-}\mu
\int_{\mathbf{S}_{2}^{(\mu)}}\mathcal{B}_{2\mathbf{k}}\mathtt{\cdot
}d\mathbf{S}\label{e25}\\
&  =\frac{e}{4\pi^{2}\lambda}\sum_{\mu=+,-}\mu\int_{\mathbf{S}_{2}^{(\mu)}%
}\nabla_{\mathbf{k}}\mathtt{\times\mathcal{A}}_{2\mathbf{k}}\mathtt{\cdot
}d\mathbf{S}\nonumber\\
&  =\frac{e}{4\pi^{2}\lambda}\sum_{\mu=+,-}\mu\oint_{\mathbf{C}_{2}^{(\mu)}%
}\mathcal{A}_{2\mathbf{k}}\mathtt{\cdot}d\mathbf{k.}\nonumber
\end{align}
We choose a symmetric form for the spin gauge potential $\mathcal{A}%
_{2\mathbf{k}}$,
\begin{equation}
\mathcal{A}_{2\mathbf{k}}=\frac{1}{2}\left(  \frac{k_{y}}{k}\text{,}%
\mathtt{-}\frac{k_{x}}{k}\right)  =k\mathbf{A}_{2\mathbf{k}}, \label{e26}%
\end{equation}
which obviously satisfies $\nabla_{\mathbf{k}}\mathtt{\times}\mathcal{A}%
_{2\mathbf{k}}$=$\mathcal{B}_{2\mathbf{k}}$. By substitution of Eq.
(\ref{e26}) in Eq. (\ref{e25}), we have
\begin{align}
\sigma_{xy}^{s0}  &  =\frac{e}{4\pi^{2}\lambda}\sum_{\mu=+,-}\mu
\oint_{\mathbf{C}_{2}^{(\mu)}}k\mathbf{A}_{2\mathbf{k}}\mathtt{\cdot
}d\mathbf{k}\label{e27}\\
&  =\frac{e}{4\pi^{2}}\sum_{\mu=+,-}\frac{\mu k_{F}^{(\mu)}}{k_{F}^{(+)}%
-k_{F}^{(-)}}\gamma_{2}^{(\mu)},\nonumber
\end{align}
where $k_{F}^{(\pm)}$ are the Fermi wave vectors for the two middle bands
$\epsilon_{2\mathbf{k}}^{\text{(}\pm\text{)}}$=$1\mathtt{-}\sqrt{3}k\pm
\sqrt{3}\lambda/2$, and we have used the fact $k_{F}^{+}\mathtt{-}k_{F}^{-}%
$=$\lambda$. Thus, we get a remarkable relationship between the conventional
SHC and Berry phases for the \textbf{K}-valley Hamiltonian. Using the chosen
middle-band eigenstates given above Eq. (\ref{e23}), it is simple to obtain
the Berry phases as $\gamma_{2}^{(+)}$=$\gamma_{2}^{(-)}$=$\pi$, leading Eq.
(\ref{e27}) to $\sigma_{xy}^{s0}$=$\frac{e}{4\pi}$, consistent again with the
numerical result in Fig. 3(b).

In summary, we have theoretically investigated the metallic spin-Hall effect
in the 2D kagom\'{e} lattice with Rashba SO coupling. When varying the Fermi
energy $\epsilon_{F}$, we have found that the conserved SHC $\sigma_{xy}^{s}$
and its two components, i.e., the conventional term $\sigma_{xy}^{s0}$ and the
spin-torque-dipole term $\sigma_{xy}^{s\tau}$, are characterized by a series
of plateaus, which is absent in the simple 2DEG system. In the whole range
$\epsilon_{F}$ varies, the two terms $\sigma_{xy}^{s0}$ and $\sigma_{\mu\nu
}^{s\tau}$ have opposite contributions. The magnitude of $\sigma_{\mu\nu
}^{s\tau}$ is twice of that of $\sigma_{\mu\nu}^{s0}$. It has been shown that
these SHC plateaus in the different regions of $\epsilon_{F}$ are closely
associated with the topologically different Fermi surfaces surrounding the
high-symmetry BZ points, i.e., the $\mathbf{\Gamma}$ and $\mathbf{K}$ points.
Thus, as has been revealed in this paper, a relationship between these SHC
plateaus and Berry phases accumulated by adiabatic motion of quasiparticles on
the Fermi surfaces can be built up, which is similar to the metallic AHE. In
particular, we have shown that compared to the case with the Fermi surface
encircling the $\mathbf{\Gamma}$ point, the amplitude of the SHC with the
Fermi surface encircling the $\mathbf{K}$ points is twice as large.
Considering the combined fact that (i) the 2D kagom\'{e} lattice is line graph
of the honeycomb structure, (ii) the Rahsba SO\ coupling and the Fermi surface
surrounding the $\mathbf{K}$ points can be easily realized in the graphene
with honeycomb structure, and (iii) the similar Berry-phase AHE has been
recently observed, we expect that the present prediction of the $\mathbf{K}%
$-valley enhanced SHE can be observed in the graphene system.

PZ was supported by NSFC under Grants No. 10604010 and No. 10534030,
 and by the National Basic Research Program of China (973 Program)
under Grant No. 2009CB929103. SL was supported by NSFC under Grants
No. 60776061 and No. 60521001.


\begin{thebibliography}{99}
\bibitem {Note}To simply clarify this feature, let us consider a usual
Anderson insulator, in which the eigenenergy states are loclized and thus the
spin displacement operator $s_{z}\mathbf{r}$ is well defined. Since the
conserved spin current $\mathcal{J}_{s}$ is given as a time derivative of the
spin displacement operator, then it is straightforward to see that the
response coefficient $\sigma_{xy}^{s}$ of $\mathcal{J}_{s}$ to a weak electric
field is zero, although both of the two components $\sigma_{xy}^{s0}$ and
$\sigma_{xy}^{s\tau}$ in $\sigma_{xy}^{s}$ can be nonzero. This subsequently
results in the conclusion that these two components have different signs.
\end{thebibliography}
\end{document}